\documentclass[preprint,showpacs,preprintnumbers,amsmath,amssymb]{revtex4}
\usepackage{graphicx}

\begin{document}
\def \ee {\varepsilon}
\thispagestyle{empty}
{\bf
Comment on ``Thermal Lifshitz force between an atom and a conductor
with a small density of carriers''
}

The application of the Lifshitz theory to describe the thermal Casimir
force in real materials leads to
 problems connected with the violation of Nernst's theorem
and contradictions with experiment \cite{9}. The Letter \cite{3}
proposes a generalization of the Lifshitz theory
 taking into account the penetration of
the static component of the fluctuating electric field into a conductor
to a depth of the Debye-H\"{u}ckel radius. We show that the proposed
generalization is thermodynamically and experimentally inconsistent.
The possible cause for this is indicated.

It was shown \cite{6}
that the proposed theory violates Nernst's theorem for the
fluctuating field in the case of dielectric plates made of
semiconductors with the concentration of charge carriers, $n$, below
critical, some semimetals, and
 solids with ionic conductivity. For these materials,
 $n$ does not go to zero when
temperature vanishes, but the conductivity $\sigma$ goes to zero due to
the vanishing mobility $\mu$. The Letter \cite{3}
 notes that
``it is difficult to estimate the number of ions which are effective
in mobility and screening'' in SiO${}_2$ glass. However,
 ionic charge carrier concentration can  be obtained by
the method \cite{7} which allows
 measuring the concentration of just those charges which
produce the screening effect. Then the $T$-dependence of $\mu$ is determined
from the $T$-dependence of $\sigma$ using $\sigma=|e|\mu n$.
Because of this, it is incorrect to transfer the $T$-dependence from
$\mu$ to $n$, as done in \cite{5}  to avoid the
violation of Nernst's theorem in \cite{3,5} for this class of
materials.

The theory of \cite{3} is also in disagreement with measurements of
the difference Casimir force $\Delta F$ between an Au sphere and a Si plate
in the presence and in the absence of laser light \cite{8}.
In Fig.~1 the dots labeled 1 show the quantity
$\Delta F^{\rm theor}-\langle\Delta F^{\rm expt}\rangle$
(for absorbed power 4.7\,mW) where $\Delta F^{\rm theor}$ was computed
using the standard Lifshitz theory with the conductivity of Si neglected
in the dark phase. In the presence of light, charge carriers were taken
into account by means of the plasma model. For dots labeled 2,
 $\Delta F^{\rm theor}$ was computed using the theory of \cite{3}.
 {}From Fig.~1 it follows that the theory of \cite{3} is
experimentally excluded at a 70\% confidence level (the opposite
conclusion obtained in \cite{5} is based
on an incorrect comparison of the experimental and theoretical results
at different confidence levels).

 According to \cite{3}, for SiO${}_2$
the relaxation time is $\tau\sim 917$ hours and ``at a such slow relaxation,
the carriers mobility can hardly be important in any experiments".
This conclusion is in conflict with the formalism of \cite{3} and
in fact favors the  prescription of \cite{9}
that for dielectrics the dc conductivity should be disregarded.
Physically, the theory of \cite{3} includes the effect of screening, i.e.,
the formation of
nonzero gradients of $n$. This situation is  out of thermal equilibrium
which is the basic applicability condition of the formalism of
\cite{3}. The violation of thermal equilibrium is the reason
why the suggested theory is experimentally and
thermodynamically inconsistent. For metals, the theory of \cite{3},
generalized in \cite{4,5}, leads  to the same results as the standard Drude
model approach. These results are in violation of the Nernst theorem for
metals with perfect crystal lattices \cite{9} and are excluded by
experiment at a 99.9\% confidence level \cite{Decca}.
Therefore it would be premature to believe that the proposed theory,
considering the static component of a fluctuating electromagnetic field
as a classical external field with nonzero magnitude, leads
to the resolution of existing problems. \hfill \\[2mm]
\noindent
B.~Geyer,${}^1$ G.~L.~Klimchitskaya,${}^1$ U.~Mohideen,${}^2$
and V.~M.~Mostepanenko${}^1$ \hfill \\
${}^1$Institute for Theoretical
Physics, Leipzig University,
D-04009, Leipzig, Germany \hfill \\
${}^2$Department of Physics and Astronomy, University of California,
Riverside, CA 92521, USA\hfill \\[2mm]
PACS numbers: 34.35.+a, 42.50.Nn, 12.20.-m

\begin{figure}
\vspace*{-4cm}
\centerline{
\includegraphics{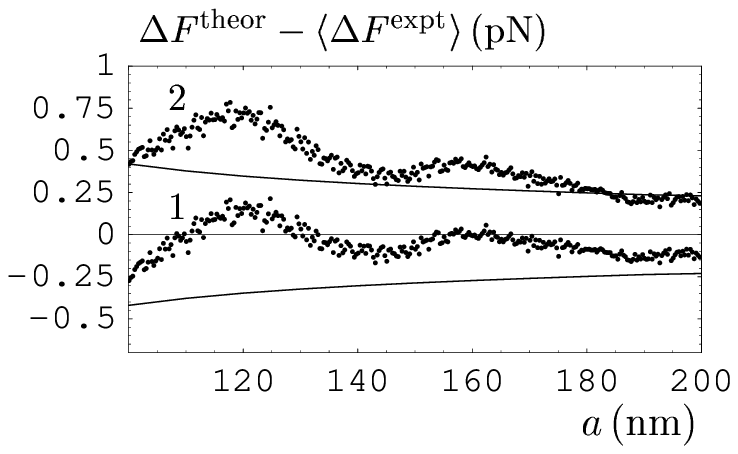}
}
\vspace*{-12cm}
\caption{Theoretical minus mean experimental differences of the
Casimir force for the Lifshitz theory (label 1) and the theory of
\cite{3} (label 2) are shown as dots versus separation. The solid lines
indicate 70\% confidence intervals including all experimental and
theoretical errors.
}
\end{figure}

\begin{thebibliography}{9}
\bibitem{9}
G.~L.~Klimchitskaya and B.~Geyer,
J. Phys. A {\bf 41}, 164032 (2008).
\bibitem{3}
L.~P.~Pitaevskii,
Phys. Rev. Lett. {\bf 101}, 163202 (2008).
\bibitem{6}
G.~L.~Klimchitskaya {\it et al.},
J. Phys. A {\bf 41}, 432001 (2008).
\bibitem{7}
M.~Tomazawa and D.-W.~Shin,
J. Non-Cryst. Sol. {\bf 241}, 140 (1998).
\bibitem{5}
V.~B.~Svetovoy, Phys. Rev. Lett. {\bf 101}, 163603 (2008).
\bibitem{8}
F.~Chen {\it et al.},
Phys. Rev. B {\bf 76}, 035338 (2007).
\bibitem{4}
D.~A.~R.~Dalvit and S.~K.~Lamoreaux,
Phys. Rev. Lett. {\bf 101}, 163203 (2008).
\bibitem{Decca}
R.~S.~Decca et al., Phys. Rev. D {\bf 75}, 077101 (2007).

\end{thebibliography}
\end{document}